\documentclass[%
 amsmath,amssymb,
 aps,prl,superscriptaddress
]{revtex4-2}
\usepackage{graphicx}% Include figure files
\usepackage{dcolumn}% Align table columns on decimal point
\usepackage{bm}% bold math
\usepackage{subfig}
\usepackage{float}
\usepackage{color}
\usepackage{subcaption}
\usepackage[colorlinks=true,
            linkcolor=blue,   
            citecolor=blue,  
            urlcolor=blue]  
            {hyperref}
\begin{document}

\preprint{APS/123-QED}

\title{Cuprate d-wave superconductivity based on Non perturbative many body theory for spin fluctuation}% Force line breaks with \\

% \thanks{A footnote to the article title}%

% 先统一放单位，再放对应作者，revtex4-2 唯一合规顺序
\author{Jianwei Gong}
\author{Junnian Xiong}
\author{Ruitao Xiao}
\affiliation{School of Physics, Peking University, Beijing 100871, China}

\author{Hui Li}
\affiliation{Institute for Advanced Study in Physics, Zhejiang University, Hangzhou 310058, China}

\author{Ziyu Li}
\author{Huaqing Huang}
\email{huanghq07@pku.edu.cn}
\author{Dingping Li}
\email{lidp@pku.edu.cn}
\affiliation{School of Physics, Peking University, Beijing 100871, China}

\thanks{*Corresponding authors}

% \collaboration{CLEO Collaboration}%\noaffiliation

\date{\today}
        
\begin{abstract}
   We employ the nonperturbative many-body spin GW method to compute the single-particle Green's function, and the covariance method to evaluate the two-body correlation functions, for the Hubbard model of superconductors. The method, benchmarked at strong coupling and low temperatures, allows simulations on lattices up to $64\times64$, thereby enabling access to the thermodynamic limit. We find the key transition temperatures---namely the N\'eel temperature and the superconducting critical temperature $T_c$---to be in quantitative agreement with experimental data. Above the $d$-wave superconducting dome, there exists a broad pseudogap region in which the normal-state excitations are suppressed. These results indicate our approach may offer a new route toward investigating strongly correlated systems, especially cuprate superconductors.
\end{abstract}

%\keywords{Suggested keywords}%Use showkeys class option if keyword
                              %display desired
\maketitle

%\tableofcontents

% ========== 删除下面这一行多余的 \documentclass ==========

% ---------- 正文开始（全部保留原 input 调用）----------

% 引言
\twocolumngrid

\noindent\textit{Introduction}---
The discovery of high-temperature superconductivity in cuprate materials by Bednorz and M\"uller in 1986~\cite{Bednorz1986} initiated one of the most profound challenges in modern physics. 
The cuprate phase diagram is characterized by a rich variety of intertwined orders \cite{fradkin2015intertwined,keimer2015}: an antiferromagnetic (AF) Mott insulating state at half-filling~\cite{Wang2020}, a pseudogap regime at intermediate doping marked by a partial suppression of low-energy electronic excitations~\cite{damascelli2003arpes,Hashimoto2014}, a `strange metal' phase featuring linear-in-temperature resistivity \cite{PhysRevB.108.134502,hussey2008transport,pnas.110.30.12235},
and a d-wave superconducting dome~\cite{RevModPhys.84.1383,RevModPhys.72.969} with critical temperatures (\(T_c\)) far beyond the McMillan limits predicted by conventional BCS theory~\cite{bardeen1957, chu1993}. The theoretical difficulty lies precisely in accounting for all these competing orders within a unified framework. Anderson's resonating valence bond theory offers a qualitative starting point, capturing d-wave pairing and the proximity to antiferromagnetism, but remains phenomenological and cannot quantitatively reproduce the full phase diagram without additional assumptions \cite{anderson1987, lee2006}. 
After four decades, the microscopic origin of high \(T_c\) superconductivity remains an unsettled problem.

It is believed that the two-dimensional single-band Hubbard model is the minimal model for describing high-temperature superconductivity in cuprate compounds~\cite{10.1098/rspa.1963.0204,qin2022hubbard,annurev:/content/journals/10.1146/annurev-conmatphys-031620-102024,PhysRevX.11.011058}. However, simple mean-field theory fails in the strongly correlated regime for this model \cite{PhysRevB.108.035139,PhysRevB.109.235149}. To overcome this problem, a variety of numerical exact methods and approximate many-body techniques have been developed. Nevertheless, each of these approaches comes with its own limitations.

Ref. \cite{Xu2025neutral} reports substantial cooling down in cold-atom quantum simulators of the Hubbard model. Such progress enables observation of previously inaccessible spin-fluctuation phases, yet temperatures remain too high to access superconductivity. Over the past five years, numerically exact studies of the Hubbard and \(t\)-\(J\) models~\cite{PhysRevB.37.3759} using the Density Matrix Renormalization Group (DMRG) method~\cite{PhysRevLett.69.2863,PhysRevB.48.10345} have made major progress. The ground-state phase diagram of the Hubbard and \(t\)-\(t'\)-\(J\) model has been mapped~\cite{Xu2024,jiang2021stripe,white2019,doi:10.1073/pnas.2109978118,PhysRevLett.127.097003,PhysRevB.106.174507,PhysRevB.108.054505,PhysRevLett.132.066002,PhysRevLett.134.116502}, and finite-temperature calculations~\cite{PhysRevLett.133.256003} have been extended to the Hubbard model~\cite{li2026fluctuatingpairdensitywave}. Calculations find \(d\)-wave superconductivity only in the electron-doped regime~\cite{PhysRevB.109.085121,PhysRevLett.129.177001}, consistent with Ref.~\cite{li2026fluctuatingpairdensitywave}. Recently, Liu et al. employed finite-size fermionic projected entangled pair states combined with variational Monte Carlo to simulate the ground state of the two-dimensional Hubbard model~\cite{r4q9-4yvj}. However, finite-size effects remain substantial due to limited system sizes~\cite{PhysRevB.84.115116}.

Determinant quantum Monte Carlo (DQMC)~\cite{PhysRevB.28.4059,PhysRevD.24.2278,PhysRevB.24.4295} suffers from a severe fermion sign problem at low temperatures and finite doping, preventing access to the superconducting ground state \cite{li2019sign, mondaini2022quantum}. 
Cellular Dynamical Mean-Field Theory (CDMFT) and Dynamical Cluster Approximation (DCA) face the same fermion sign problem, which limits the achievable cluster sizes and prevents simulations at sufficiently low temperatures, as reviewed in Refs.~\cite{georges1996,RevModPhys.83.349,maier2005,doi:10.7566/JPSJ.92.092001} and references therein. In practice, the methods cannot maintain full Ward identities (for arbitrary $q$)~\cite{RevModPhys.90.025003}, which breaks the conservation laws and introduces systematic uncertainties in transport coefficients and response functions. The 
GW and FLEX approximations, based on the Luttinger-Ward functional, are widely employed because they obey conservation laws (particle number, momentum, and energy) ~\cite{baym1962, rohringer2018}.
However, these theories were thought to break down in the strong-coupling regime (\(U/t \gg 1\)) relevant to cuprates. Even in regimes where they are applicable, such approximations do not guarantee the correct normalization of key fluctuation spectra—specifically, 
the \(f\)-sum rule~\cite{Ortloff2007,PhysRevB.108.075144}. This raises fundamental concerns about their quantitative reliability for describing the collective excitations.

% \textbf{Numerically exact methods} represent the gold standard for benchmarking. \textbf{Quantum Monte Carlo (QMC)} simulations of the Hubbard model are free from the fermionic sign problem only at half-filling. For the doped two-dimensional (2D) case of interest, 
% the sign problem becomes exponentially severe at low temperatures\cite{li2019sign,mondaini2022quantum}, thereby precluding access to the superconducting ground state. \textbf{Tensor network} methods, such as the density matrix renormalization group (DMRG), have made significant progress, revealing stripe order and pairing tendencies in quasi-1D ladders and small 2D cylinders~\cite{white2019, jiang2021stripe}. 
% However, extending these methods to the true 2D thermodynamic limit at finite doping and low temperatures remains an immense computational challenge, and it is unclear how to reliably extrapolate cylinder results to the full 2D limit.

Cuprates are not superconductors emerging from a featureless normal state; instead, antiferromagnetism, stripe order, and the pseudogap all compete with superconductivity. A successful theory must therefore account for the entire phase diagram, not just the superconducting dome. To provide a reliable description of strongly correlated electron systems and enable quantitative comparisons with experimental measurements, such a theory should satisfy several essential requirements: (i) non-perturbative treatment of strong correlations due to large on-site $U$; (ii) systematic inclusion of long-range spatial and temporal fluctuations; (iii) enforcement of conservation laws and sum rules (e.g., via Ward identities); (iv) computational tractability for large two-dimensional systems at low temperatures; and (v) the ability to calculate both single-particle and two-particle observables, enabling quantitative predictions for photoemission, transport, and scattering experiments \cite{Hashimoto2014,damascelli2003arpes, scalapino2012, hussey2008transport,pnas.110.30.12235}. No existing theoretical method simultaneously satisfies all these requirements; consequently, a reliable theory of high-temperature superconductivity in cuprates remains out of reach.

In this letter, we propose such a framework: a self-consistent, symmetry-preserving, non-perturbative approximation scheme, covariant spin GW(variant GW different from the often used GW~\cite{baym1962, rohringer2018}) approximation theory~\cite[and references therein]{10.21468/SciPostPhys.20.4.101}. 
By construction, the approach satisfies all Ward identities and conserves particle number, momentum, and energy. The framework was benchmarked against DQMC results for the half-filled Hubbard model with \(U/t \le 8\) and $\beta \le 16$(for size$=16\times 16,\beta \geq 20$, the time of DQMC is costly)~\cite{10.21468/SciPostPhys.20.4.101}, demonstrating its credibility for strongly correlated systems at half-filling. It provides a systematic pathway to compute single-particle spectral functions and two-particle fluctuation spectra away from half-filling, enabling direct comparison with ARPES, Raman, and neutron scattering experiments. The phase diagrams and the solution of the GW equations in different phase regimes will be presented in this work.

% 理论/公式部分
\noindent\textit{Model and methods}--- 
We consider the 2D Hubbard model with periodic boundary conditions on a square lattice 
\begin{equation}
    \hat H = -\sum_{ij}\sum_{\sigma=\uparrow,\downarrow}t_{ij}
    \hat c^\dagger_{i\sigma}\hat c_{j\sigma} + U\sum_{i}\hat n_{i\uparrow}\hat n_{i\downarrow}
    -\mu\sum_{i\sigma}\hat n_{i\sigma}.\label{eq:Hamiltonian}
\end{equation}

Here $\hat c^\dagger_{i\sigma}$ ($\hat c_{i\sigma}$) is the creation (annihilation) operator 
for an electron with spin $\sigma$ at lattice site i. 
$\hat n_{i\sigma}=\hat c^\dagger_{i\sigma}\hat c_{i\sigma}$ 
denotes the spin-resolved density operator. 
The hopping amplitude $t_{ij}$ is denoted such that 
$t$ ($t'$, $t''$) corresponds to nearest- (next-nearest-, next-next-nearest-) neighbor hopping. 
We take the on-site Coulomb repulsion $U>0$ as the 
interaction strength, with chemical potential $\mu$, 
and set all energies in units of $t=1$ with $\hbar = 1$.

Based on the coherent-state path-integral quantization, we investigate the competing instabilities of the two-dimensional Hubbard model.
In our numerical framework, the spin-GW approach is employed to compute the single-particle Green’s function, while the covariant correction formalism is utilized to evaluate two-body correlation functions.Detailed derivations and numerical implementations are provided in the Supplemental Material \cite{gong2026supplement}.
This scheme strictly satisfies the fluctuation-dissipation theorem and Ward identity.

The $d$-wave superconducting order parameter is defined as~\cite{qin2022hubbard,lee2006}:
\begin{equation}
\Delta^d_i = \frac{1}{2}\left[
c^\dagger_{i\downarrow} \big(c^\dagger_{i+e_x\uparrow} + c^\dagger_{i-e_x\uparrow} 
- c^\dagger_{i+e_y\uparrow} - c^\dagger_{i-e_y\uparrow}\big)
\right],
\end{equation}

To establish the phase diagram, we systematically calculate three characteristic two-body susceptibilities: 
the $d$-wave pairing susceptibility $\chi^d$, the spin susceptibility $\chi^s$, and the charge-density-wave susceptibility $\chi^c$.
The low-temperature divergence of $\chi^d$ indicates the instability toward $d$-wave superconductivity.
The divergence of the spin susceptibility $\chi^s$ at the antiferromagnetic wave vector $Q^\text{AF}$ defines the AF branch instability from high temperature.
Meanwhile, the divergence of the charge susceptibility $\chi^c$ at finite momentum $Q^\text{CDW}$ corresponds to the emergence of charge-density-wave order.
The full numerical phase diagram is constructed by tracing these instability boundaries over varying particle density and interaction strength.

\begin{figure*}[htbp]
\centering
\includegraphics[width=\textwidth]{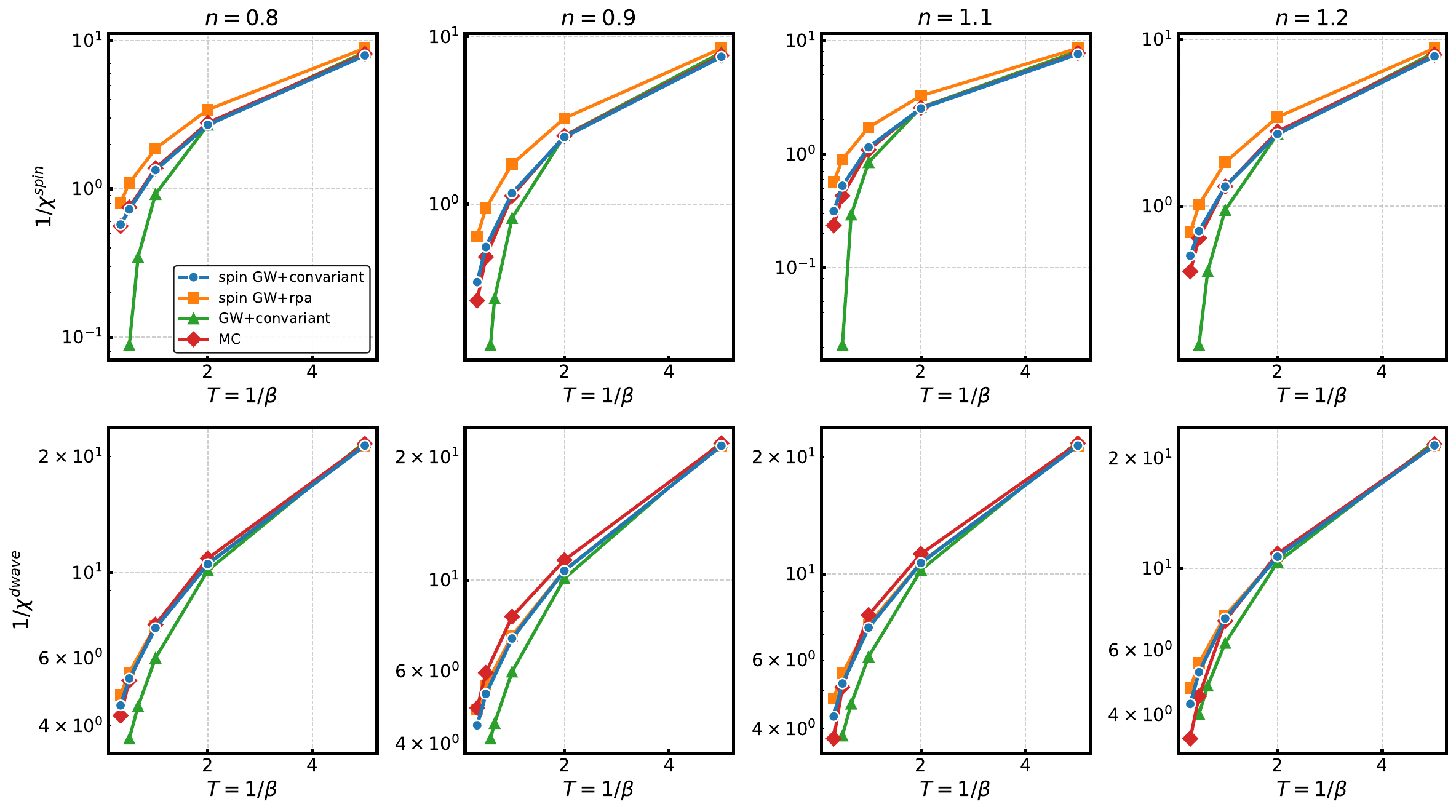}
\caption{Numerical results of spin and $d$-wave susceptibilities for $U=7$, $t'=-0.25$ and lattice size $N=8\times8$, obtained from four different numerical approaches at four distinct fillings (from left to right).
The four methods include spin GW+convariant, GW+convariant, spin GW+random-phase approximation (RPA), and DQMC.
Each panel displays the inverse susceptibility $1/\chi$ as a function of temperature $T$.
Spin susceptibility $\chi^s$ is shown in the upper row, and $d$-wave pairing susceptibility $\chi^d$ is plotted in the lower row.
The filling $n$ for each column is indicated above each subfigure.}
\label{fig:fig0}
\end{figure*}

\noindent\textit{Benchmark test in doped regime}--- 
Due to the severe sign problem encountered by DQMC in the low-temperature doped regime, we compare the results from three many-body approaches with DQMC data within the high-temperature window($T>1/3$). As illustrated in Fig.~\ref{fig:fig0}, we compute the spin and d-wave susceptibilities for various fillings. Here, spin GW+RPA corresponds to the approximation retaining only the bare vertex \(\gamma\) and Hartree vertex contribution \(\Gamma_{\text{H}}\)~\cite{gong2026supplement}. GW+covariant denotes the often used GW formalism adopted in this work, which incorporates fluctuations in the density channel while omitting spin fluctuations. For DQMC simulations, we utilize an existing implementation~\cite{SciPostPhysCodeb.29-r0.3}. The benchmark in Fig.~\ref{fig:fig0} demonstrates the advantages of spin GW+covariant and supports the validity of our results extended to low temperatures.

\begin{figure*}[htbp]
\centering
\includegraphics[width=\textwidth]{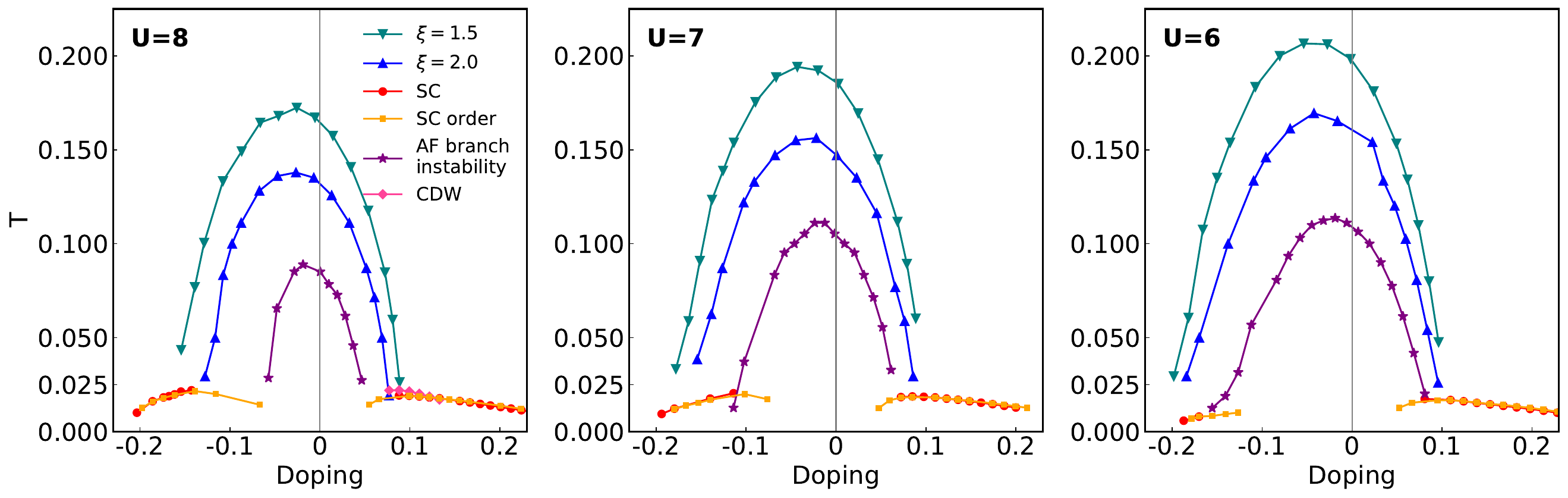}
\caption{Numerical phase diagram of the Hubbard model for different interaction strengths $U$ with fixed next-nearest-neighbor hopping $t'=-0.25$ and lattice size $32\times32$.
$\boldsymbol{\text{SC order}}$: stable boundary with $\Delta^d\neq0$;
$\boldsymbol{\text{SC}}$: divergence boundary of $\chi^d(0,0,0)$;
$\boldsymbol{\xi=1.5}$ and $\boldsymbol{\xi=2.0}$: contours of the spin correlation length $\xi$ equal to 1.5 and 2, respectively; the region enclosed between these two contours is identified as the pseudogap crossover regime;
$\boldsymbol{\text{AF branch instability}}$: crossover from weak to strong correlated AF fluctuations, rather than a true AF phase transition;
$\boldsymbol{\text{CDW}}$: divergence of $\chi^c(Q^\text{CDW},0)$.}
\label{fig:fig1}
\end{figure*}

\noindent\textit{Phase diagram}--- 
Figure.~\ref{fig:fig1} presents the phase diagram of the two-dimensional $t$-$t'$-$U$ Hubbard model obtained using the GW+covariance method. The diagram covers both electron- and hole-doped regimes (left and right sides, respectively). For $U=6-8$, as the temperature decreases from high to low, $\chi^{s}(\boldsymbol{Q}^{\text{AF}},i\omega_n)$ \cite{gong2026supplement} diverges at $i\omega_n=0$ when the carrier density lies within a moderate range. Our spin GW+covariance calculations reveal a divergence which indicates high-temperature branch instalbility of $\chi^{\text{spin}}$. It is not a genuine phase transition, but a pseudo-critical crossover. This is precisely the behavior identified in the exact O($\infty$) solution of the 2D Heisenberg model~\cite{Butera1989ComplexTS} and in tensor-network studies~\cite{SciPostPhys.11.5.098}, where apparent divergences arise without finite-$T$ criticality. Preliminary results from our post-GW calculations, combined with a linear extrapolation of the two lowest positive Matsubara frequencies~\cite{PhysRevX.8.021048}, indicate that the pseudogap onset at the antinodal point occurs when the spin correlation length \(\xi\) reaches approximately 1.75. This criterion is qualitatively consistent with the empirical threshold reported in Ref.~\cite{7v3y-tbbr}. However, the post-GW approach remains computationally demanding~\cite{Li2026postgw}, and a full implementation is deferred to future work. For the present study, we therefore adopt a pragmatic definition of the pseudogap crossover regime by drawing contours of \(\xi=1.5\) and \(\xi=2.0\), with the region between them identified as the pseudogap crossover.

The phase boundary of $d$-wave superconductivity is cross-validated using two independent methods. The first method identifies the divergence positions of $\chi^{d}(\boldsymbol{Q}=0,i\omega_n=0)$, while the second determines the boundary where stable solutions with a non-zero order parameter emerge. The boundaries obtained from these two methods agree well in regions far from the AF branch instability. 

For the phase diagram at $U=8$, we also evaluate charge density instabilities. The density correlation function is found to diverge near the wave vector $\boldsymbol{Q}^{\text{CDW}}=(\pi/4,0)$, which defines the phase boundary of the charge density wave (CDW) state. For $U=6$ and $U=7$, the superconducting long-range order emerges at a higher temperature than the CDW order, indicating a competition between CDW and $d$-wave superconductivity. Moreover, no CDW order is observed in the electron-doped region for these interaction strengths.

\begin{figure}[ht]
\centering
\includegraphics[width=0.4\textwidth]{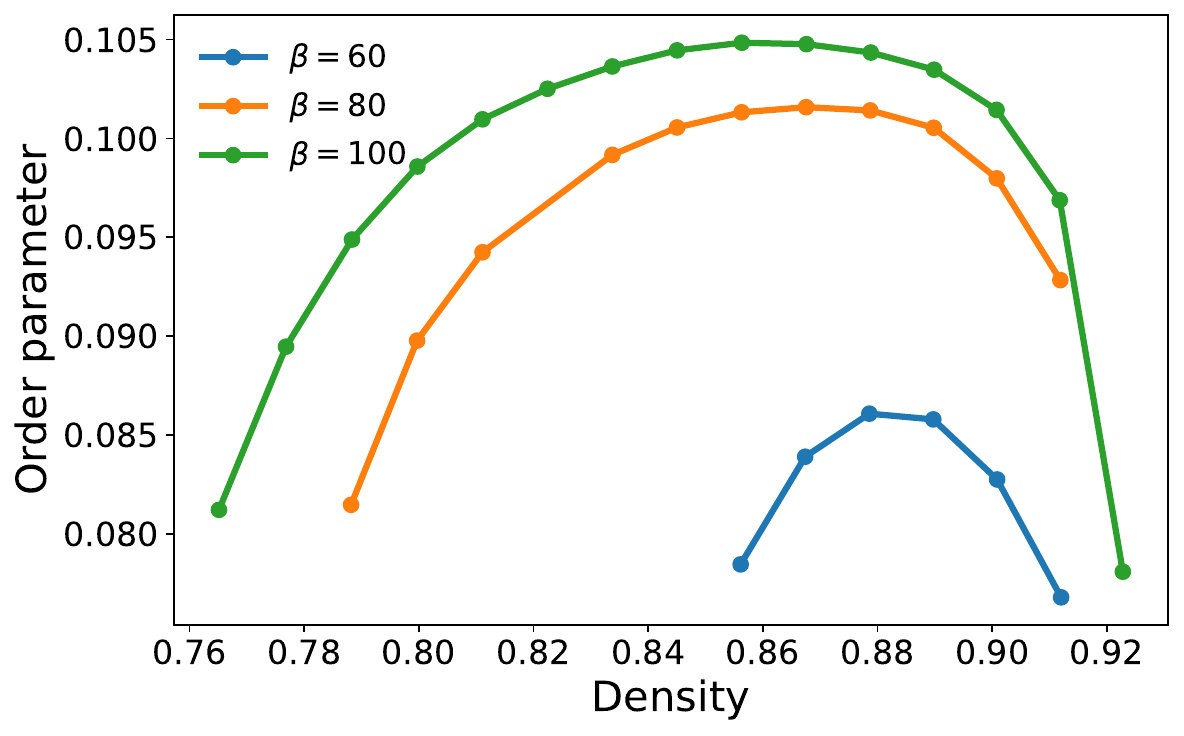}
\caption{Magnitude of the $d$-wave superconducting order parameter $|\Delta^d|$ as a function of doping density in the hole-doped region for $t'=-0.25$, $U=8$, and lattice size $N=32\times32$ at various temperatures.}
\label{fig:fig2}
\end{figure}

\noindent\textit{Dwave superconducting solution}--- 
Figure.~\ref{fig:fig2} shows the variation of the $d$-wave order parameter with doping at several temperatures in the superconducting region. The order parameter reaches a maximum at the optimal doping level, which is consistent with the experimentally observed dome-shaped $T_c$ curve. As the doping deviates from this optimal value, the order parameter decreases monotonically, reflecting the weakening of pairing correlations in both underdoped and overdoped regimes. Notably, the maximum of $|\Delta^d|$ also decreases with increasing temperature, as expected from the suppression of superconductivity by thermal fluctuations. These results further support the validity of the computed phase boundary and highlight the close connection between the order parameter magnitude and the superconducting transition temperature. Furthermore, the Supplemental Material~\cite{gong2026supplement} illustrates the real-space distribution of the anomalous Green's function for the $d$-wave superconducting solution. Apart from the dominant nearest-neighbor $d_{x^2-y^2}$ pairing component, additional $d_{x^2-y^2}$-wave components with different pairing lengths emerge.

\begin{figure}[ht]
\centering
\includegraphics[width=0.4\textwidth]{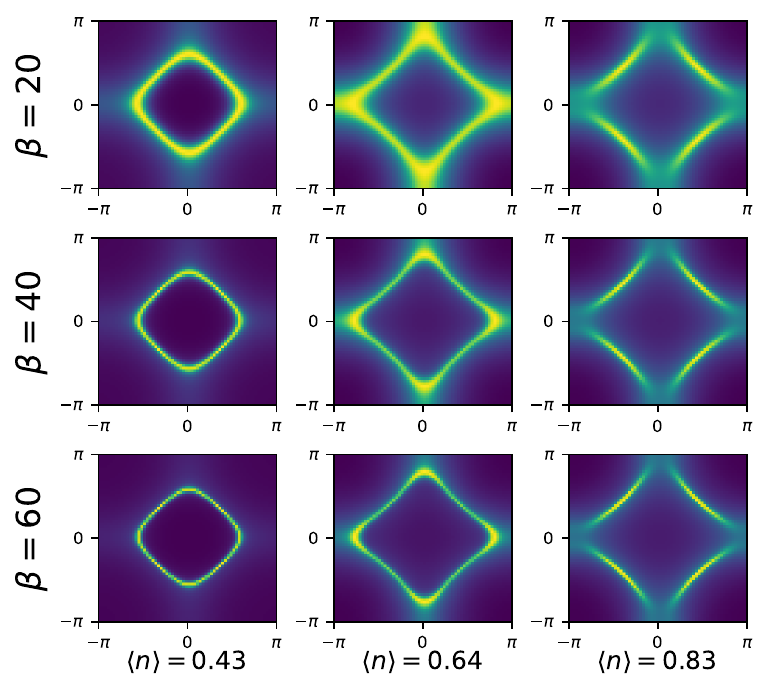}
\caption{Color plots of \(\partial n_k/\partial \mu\) for the Hubbard model with \(U=8\) and \(t'=-0.25\) on a \(64\times64\) square lattice. The three columns correspond to different hole-doping levels, and the three rows correspond to different inverse temperatures \(\beta\) (from top to bottom: \(\beta=20, 40, 60\)).}
\label{fig:fig3}
\end{figure}

\noindent\textit{Fermi surface topology and Fermi arcs}--- 
Using the peak positions of \(\partial n_k/\partial \mu\) to identify the Fermi surface~\cite{PhysRevB.107.165154}, we have indirectly mapped out the evolution of the Fermi surface topology in the hole-doped regime of the Hubbard model with \(U=8\) and \(t'=-0.25\) (Fig.~\ref{fig:fig3}). This approach circumvents the need for direct analytic continuation of the Green's function, and its validity has been benchmarked~\cite{PhysRevB.107.165154}. As hole doping decreases, the Fermi surface undergoes a systematic evolution driven by a Lifshitz transition~\cite{PhysRevLett.114.147001}, from convex(electron-like) to concave(hole-like), and eventually to disconnected Fermi arcs~\cite{PhysRevB.74.125110,PhysRevB.106.245132}. These three regimes correspond to the Fermi liquid, strange metal, and pseudogap regions, respectively, with the latter commonly associated with the emergence of Fermi arcs. The crossover boundaries determined by this method are in qualitative agreement with experimental observations, demonstrating the reliability of our self-consistent covariant $GW$ scheme.

Overall, the phase diagram and the doping-dependent order parameter presented here provide a comprehensive picture of the competing orders in the Hubbard model and offer a benchmark for future theoretical and experimental studies of cuprate superconductors.
% 结论
\noindent\textit{Discussion and outlooks}---
In summary, the method used in this paper meets the basic five criterions raised in the introduction: in particular FDT, WTI and size which is big enough and computational achievable for thermal dynamical limit.

Unlike GW for negative U Hubbard model that includes the pairing channel \cite{973p-gphs}, where the fluctuation of the superconducting order parameter field is included, the present GW used in this work only incorporates spin fluctuations while superconducting order parameter field fluctuation is ignored, the resulting superconducting transition temperature for d-wave superconductivity is expected to be BCS-like or mean field critical temperature  in contrast to negative U \cite{973p-gphs}, which is indeed confirmed by finite-size scaling of superconducting phase transition temperature in this letter~\cite{gong2026supplement}.
Mean-field theory doesn't include spin order parameter field  fluctuation,and the methods failed to yield d-wave superconducting solution \cite{PhysRevB.108.035139,PhysRevB.109.235149}. The results confirm that spin fluctuations contribute the dominant pairing mechanism.
The real superconducting phase transition temperature will be renormalized and lower than the mean field critical temperature obtained in the present work. Using the Ginzburg--Landau--Lawrence--Doniach model~\cite{RevModPhys.82.109,jiang2014thermalfluct} and taking into account fluctuations of the order parameter field, the superconducting transition temperature \(T_c\) of infinite-layer superconductors is renormalized downward, becoming lower than the mean-field critical temperature~\cite{jiang2014thermalfluct}. For example, the critical temperature of YBCO or BSCCO is about 10\% lower than the mean-field value, as demonstrated in Ref.~\cite{jiang2014thermalfluct}. A single layer of Bi$_2$Sr$_2$CuO$_{6+\delta}$ (Bi-2201), which comprises only one CuO$_2$ plane, has been fabricated and found to be superconducting in the doped regime with an optimal \(T_c\approx 35\)~K~\cite{Luo2026}. The existence of such an experimentally realized purely two-dimensional superconductor is consistent with the results of the two-dimensional Hubbard model. To go beyond the mean-field critical temperature, one needs to include diagrams beyond the GW approximation~\cite{PhysRevB.97.014528}.

Our numerical results reproduce key features of the experimental phase diagram of cuprate high-temperature superconductors. The phase diagram we obtained is qualitatively similar to those from CDMFT and DCA~\cite{M.Jarrell_2001,cpl_42_8_080711,PhysRevB.108.075163,PhysRevLett.110.216405}.The results show that $T_c$ peaks at $U=7$--$8$, with the maximum $T_c\approx 0.02t$ where t is the nearest-neighbor hopping amplitude. The amplitude can be fitted by comparing the results of DCA or CDMFT \cite{M.Jarrell_2001,cpl_42_8_080711,PhysRevB.108.075163,PhysRevLett.110.216405} (or other approximation method) to experimental data and Ab initio calculation \cite{PhysRevB.106.235150,PhysRevB.98.134501,PhysRevB.99.245155},
and its value lies between  0.35 eV and  0.54 eV, so the maximum $T_c$ falls in the ranges $80-125$K. 
The superconducting critical temperature is in the range of experimental observations for cuprate superconductors \cite{pnas.110.30.12235}.

Superconductivity in the electron-doped regime is suppressed by antiferromagnetism for $U=6$, whereas $T_c$ in the hole-doped region remains nearly unchanged.

There is CDW order instability with a transition temperature close to that of $d$-wave superconductivity in the underdoped regime for $U=8, t'=-0.25$, implying intricate competition between the CDW and superconducting phases. There are experimental evidences for the existences of the intertwined domains of stripe and superconducting orders in the underdoped regime \cite{Wen2019}. The numerical evidence such as DMRG calculations, which confirm stripe--$d$-wave superconducting coexistence in the ground state ~\cite{white2019,jiang2021stripe,Xu2024,PhysRevLett.133.256003,PhysRevLett.132.066002}.

The covariant spin GW approach employed in this work offers a new computational route to the cuprate phase diagram, but it has not yet addressed single‑particle spectral functions such as the superconducting and pseudogaps. The post‑GW method extends this framework directly: its screened potential is constructed from the covariance spin correlations, 
and it has been benchmarked against quantum Monte Carlo in both the half‑filled 2D repulsive Hubbard model (where it reproduces the pseudogap) and the 3D attractive Hubbard model (where it yields an accurate superconducting gap) \cite{Li2026postgw,973p-gphs}.
Applying post‑GW at low temperatures will thus provide spectral functions directly comparable with ARPES experiments, offering a systematic route to resolving the gap problem in cuprates.

% ---------- 附录（同样用小标题，保留 input）----------
% 原先的 \appendix 命令已删除，改用小标题区分
% \onecolumngrid
% \section*{Supplemental Material}

% \subsection{Phase diagram at other parameter values}
% \input{appendix/appendixA.tex}

% \subsection{2-Nambu spin-GW equation in the paramagnetic state}
% \input{appendix/appendixB.tex}

% \subsection{Covariance Calculation in the normal phase and superconducting phase}
% \input{appendix/appendixC.tex}

% \subsection{Details of the finite-size scaling analysis}
% \input{appendix/appendixD.tex}

% 如果附录E需要，继续
% \subsection{Some other topic}
% \input{appendix/appendixE.tex}

% 如果不需要附录E，保持注释状态即可
% \paragraph*{---\textit{Supplementary Material S5}}
% \input{appendix/appendixE.tex}

% ---------- 致谢 ----------
% \clearpage
\begin{acknowledgments}
Authors are very grateful to B. Rosenstein, Shiping Feng, Qingdong Jiang, Kui Jing, Qiaoyi Li, Wei Li, Haiqin Lin, Tianxin Ma, Mingpu Qing, Xinguo Ren, Yingze Su, Zhipeng Sun, Yigui Yao, and Xiaotian Zhang for valuable discussions.
This work is supported by the National Natural Science
Foundation of China (Grant No.12174006 of Prof. Li's fund) 
and the High-performance Computing Platform of Peking University. H.H. acknowledges 
the support of the National Key R\&D Program of China (No. 2021YFA1401600), 
and the National Natural Science Foundation of China (Grants No. 12074006 and 12474056).
\end{acknowledgments}

\nocite{*}
\bibliography{apssamp}

\onecolumngrid
%\newpage

%\pagenumbering{arabic}
%\setcounter{page}{1}
\setcounter{equation}{0}
\setcounter{figure}{0}
\setcounter{table}{0}
\setcounter{section}{0}
\renewcommand{\thefigure}{S\arabic{figure}}
\renewcommand{\thetable}{S\arabic{table}}
\renewcommand{\theequation}{S\arabic{equation}}

\begin{center}

\textbf{Supplemental Material:Cuprate d-wave superconductivity based on nonperturbative many-body theory for spin fluctuation}
\end{center}

\section*{Introduction to Supplemental Material}
This document provides additional technical details and supporting results for the main text.
It contains the following five appendices.

% ----- 附录 A -----
\section{Appendix A: Phase diagram at other parameter values}

\begin{figure*}[htb]
  \centering
  \includegraphics[width=0.95\textwidth]{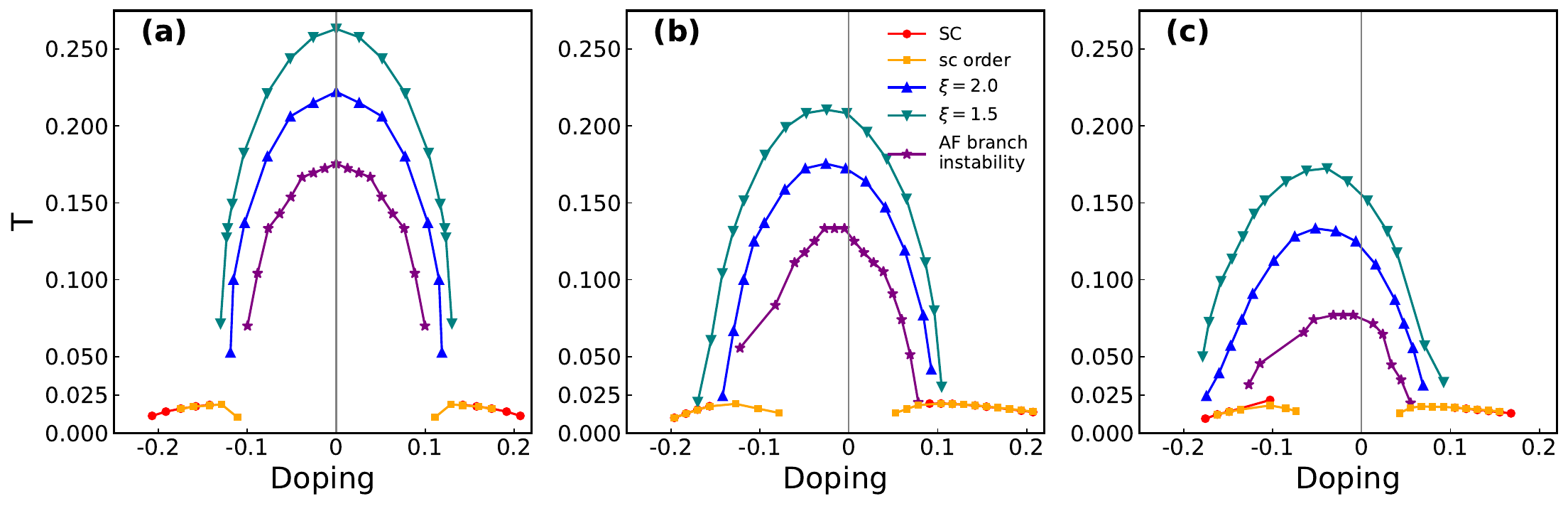}
  \caption{Phase diagrams of the $t$-$t'$-$U$ Hubbard model on a 32$\times$32 lattice for other parameters. (a) $U=6, t'=0$, (b) $U=7, t'=-0.2$, (c) $U=7, t'=-0.3$.
  $\boldsymbol{\text{SC order}}$: stable boundary with $\Delta^d\neq0$;
  $\boldsymbol{\text{SC}}$: divergence boundary of $\chi^d(0,0,0)$;
  $\boldsymbol{\xi=1.5}$ and $\boldsymbol{\xi=2.0}$: contours of the spin correlation length $\xi$ equal to 1.5 and 2, respectively;
  $\boldsymbol{\text{AF branch instability}}$: crossover of AF fluctuations (not a true phase transition).}
  \label{fig:Figure4}
\end{figure*}

Figure.~\ref{fig:Figure4} shows the phase diagrams for $t' = -0.2$ and $-0.3$. As $|t'|$ decreases, the AF region shifts toward the hole-doped region, and the phase diagrams on both sides of the doping region gradually become consistent.When $t' = 0$, we present the results for $U = 6$ in Figure.~\ref{fig:Figure4}. Due to the particle-hole symmetry, the phase diagram is symmetric between electron doping and hole doping, which indicates that even without considering next-nearest-neighbor hopping, a stable superconducting long-range order exists.
%\input{appendix/appendixA.tex}
% \section{Phase diagram at other parameter values}

\section{Appendix B: Spatial Distribution of Anomalous Green’s Function in Real Space}
The superconducting solution of Eq.~(\ref{eq:gw_momentum}) in the low-temperature regime is difficult to converge with random initial iterations. Therefore, we need to introduce a d-wave source term $h(\cos k_x - \cos k_y)\sigma_x$, and then slowly ramp $h$ down to zero, eventually obtaining a stable d-wave superconducting solution.

We also focus on the distribution of the anomalous Green's function $F(\mathbf{r},\tau=0)$ in real space.

We construct symmetry-adapted basis functions for the $d_{x^2-y^2}$ pairing channel via the group projection operator technique, which relies entirely on the character table of the $D_4$ point group summarized in Table \ref{tab:char-D4}.
For any trial function $f$, the basis function transforming under the $B_1$ irreducible representation ($d_{x^2-y^2}$ symmetry) is given by~\cite{PhysRevB.47.5977}
\begin{equation}
f^{B_{1}}=\frac{d_{B_{1}}}{|G|} \sum_{g \in G} \chi^{B_{1}}(g) \hat{O}_{g} f,
\label{eq:proj-op}
\end{equation}
where $|G|=8$ is the total number of symmetry operations in $D_4$, $d_{B_1}=1$ is the dimension of the one-dimensional $B_1$ representation, $\chi^{B_1}(g)$ denotes the character of each group element $g$, and $\hat{O}_g$ represents the spatial symmetry transformation operator.
The characters $\chi^{B_1}(g)$ used in the summation are directly extracted from the row of $B_1\;(d_{x^2-y^2})$ in the $D_4$ character table.

\begin{figure}[htb]
  \centering
  \includegraphics[width=0.60\textwidth]{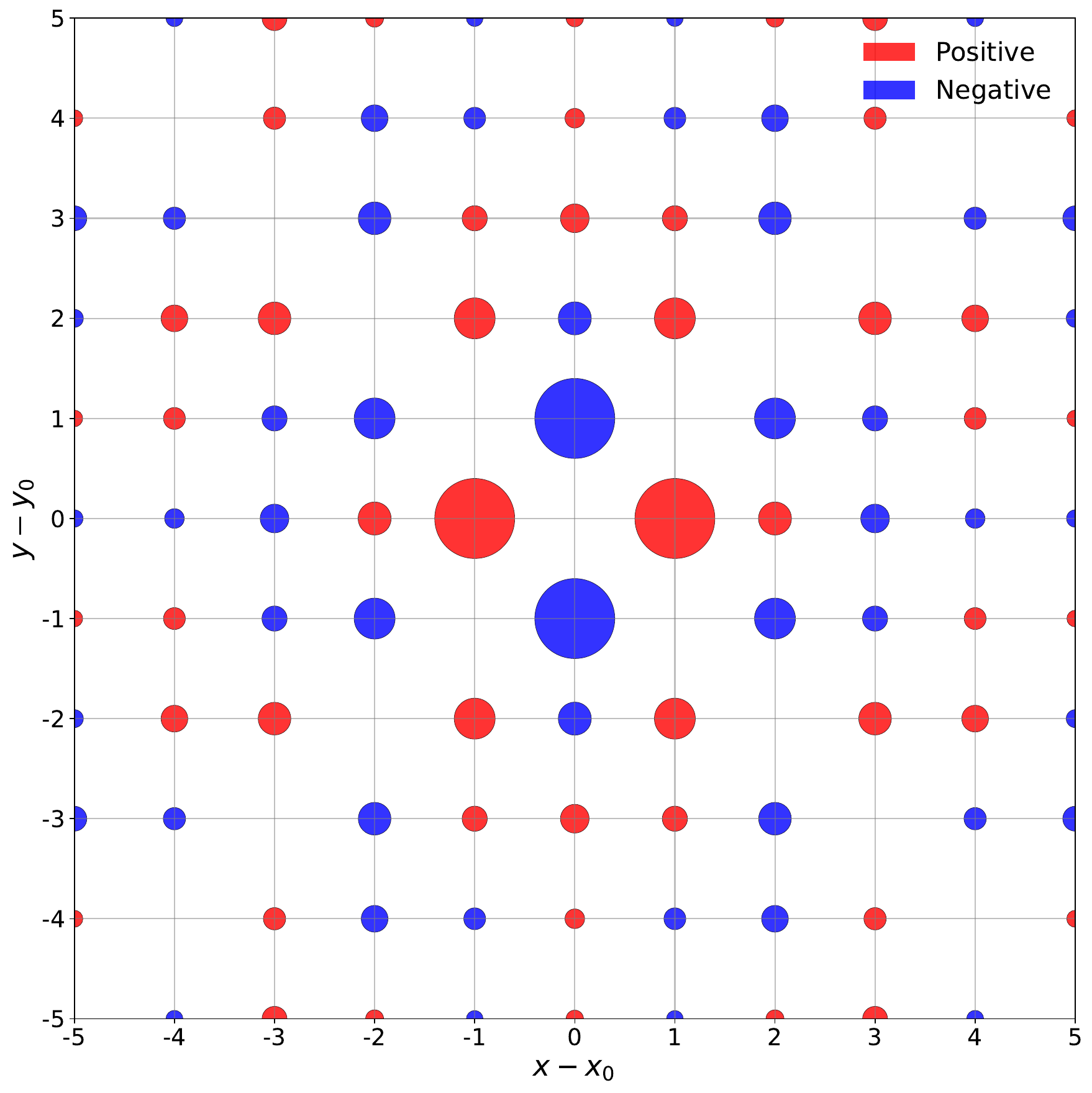}
  \caption{Real-space distribution of the pairing order parameter $F(\mathbf{r},\tau=0)$ under parameters $U=8$, $t'=-0.25$, $N=32\times32$, $T=0.01$, $T_c\approx0.018$, and electron filling $\langle n\rangle=0.879$. The area of each circle centered at lattice site $\boldsymbol{r}=(x,y)$ is proportional to the magnitude of $F(\boldsymbol{r})$. }
  \label{fig:figure5}
\end{figure}

We take plane-wave trial functions $f=\cos(\boldsymbol{k}\cdot\boldsymbol{r})$ defined on square lattice separations $(m,n)$. Substituting the eight group operations and the corresponding $B_1$ characters from Table \ref{tab:char-D4} into Eq.~\eqref{eq:proj-op}, we derive the momentum-space basis function $f_{m,n}^{d_{x^2-y^2}}(k_x,k_y)$:
\begin{align}
f_{m, n}^{d_{x^{2}-y^{2}}}\left(k_{x}, k_{y}\right)
=&\frac{1}{8}\Big[\cos \left(m k_{x}+n k_{y}\right)-\cos \left(-n k_{x}+m k_{y}\right)-\cos \left(n k_{x}-m k_{y}\right)+\cos \left(-m k_{x}-n k_{y}\right)\nonumber\\
&+\cos \left(m k_{x}-n k_{y}\right)+\cos \left(-m k_{x}+n k_{y}\right)-\cos \left(n k_{x}+m k_{y}\right)-\cos \left(m k_{x}-n k_{y}\right)\Big]\nonumber\\
=&\frac{1}{4}\left[\cos \left(m k_{x}+n k_{y}\right)+\cos \left(m k_{x}-n k_{y}\right)-\cos \left(n k_{x}+m k_{y}\right)-\cos \left(n k_{x}-m k_{y}\right)\right].
\label{eq:f-kspace}
\end{align}

A Fourier transform to real space yields the lattice delta-function form of the basis:
\begin{align}
f_{m, n}^{d_{x^{2}-y^{2}}}(x, y)=\frac{1}{4}\Big[
&\delta_{(x,y);(m,n)}+\delta_{(x,y);(m,-n)}+\delta_{(x,y);(-m,n)}+\delta_{(x,y);(-m,-n)}\nonumber\\
&-\delta_{(x,y);(n,m)}-\delta_{(x,y);(n,-m)}-\delta_{(x,y);(-n,m)}-\delta_{(x,y);(-n,-m)}
\Big].
\label{eq:f-rspace}
\end{align}
This basis $f_{m,n}^{d_{x^2-y^2}}$ has a key symmetry property: $f_{m,n}^{d_{x^2-y^2}}\equiv0$ for $m=n$, caused by complete cancellation of positive and negative weights along lattice diagonals, consistent with the alternating signs of $B_1$ characters under diagonal reflection operations.
The full $d_{x^2-y^2}$ pairing order parameter is assembled by linear superposition over all lattice vectors weighted by these symmetry basis functions:
\begin{equation}
\Delta^{d_{x^{2}-y^{2}}}(x, y)=\sum_{m, n} \Delta^{m, n} f_{m, n}^{d_{x^{2}-y^{2}}}(x, y).
\label{eq:delta-dx2y2}
\end{equation}

\begin{table}[htbp]
  \centering
  \renewcommand{\arraystretch}{1.4}
  \setlength{\tabcolsep}{8pt}
  \begin{tabular}{c|ccccc}
    $D_4$ & $E$ & $2C_4$ & $C_2$ & $2C_2'$ & $2C_2''$ \\
    \hline
    $A_1$ & $1$ & $1$ & $1$ & $1$ & $1$ \\
    $A_2$ & $1$ & $1$ & $1$ & $-1$ & $-1$ \\
    $B_1\;(d_{x^2-y^2})$ & $1$ & $-1$ & $1$ & $1$ & $-1$ \\
    $B_2$ & $1$ & $-1$ & $1$ & $-1$ & $1$ \\
    $E$ & $2$ & $0$ & $-2$ & $0$ & $0$ \\
  \end{tabular}
  \caption{Character table of the $D_4$ point group. The row labeled $B_1\;(d_{x^2-y^2})$ provides the required characters $\chi^{B_1}(g)$ for constructing the $d_{x^2-y^2}$ symmetry-adapted basis via projection operators.}
  \label{tab:char-D4}
\end{table}

\begin{table}[h]
\centering
\renewcommand{\arraystretch}{1.5}
\setlength{\tabcolsep}{12pt}
\begin{tabular}{c|ccccc}
\hline
$n \backslash m$ & 0 & 1 & 2 & 3 & 4 \\ \hline
0 & -0.000000 & -0.013043 & -0.002237 & 0.001689 & 0.000792 \\
1 & 0.013043 & 0.000000 & 0.003424 & 0.001299 & -0.000982 \\
2 & 0.002237 & -0.003424 & -0.000000 & -0.002186 & -0.001470 \\
3 & -0.001689 & -0.001299 & 0.002186 & 0.000000 & 0.001021 \\
4 & -0.000792 & 0.000982 & 0.001470 & -0.001021 & -0.000000 \\
\hline
\end{tabular}
\caption{Real-space order parameter magnitude $|\Delta|^{d_{x^2-y^2}}(m,n)$.}
\label{tab:order}
\end{table}

As shown in Fig.~\ref{fig:figure5}, the real-space pairing field \(F(\boldsymbol{r},\tau=0)\) exhibits a symmetry that closely matches the \(f_{m,n}^{d_{x^2-y^2}}\) basis derived from the \(D_4\) character table. Specifically, the \(B_1\) (\(d_{x^2-y^2}\)) pairing component dominates, whereas no \(B_2\) (\(d_{xy}\)) signal is resolved---consistent with the opposite parity of their character rows under diagonal two-fold rotations. Importantly, Significant pairing amplitudes are also observed at extended distances, notably at \(2\mathbf{e}_x\), \(2\mathbf{e}_x+\mathbf{e}_y\), and \(3\mathbf{e}_x+2\mathbf{e}_y\). Specifically, the pairing amplitude at \(2\mathbf{e}_x+\mathbf{e}_y\) is approximately 1.5 times that at \(2\mathbf{e}_x\). These extended components, which involve higher-order \((m,n)\) basis terms, contribute substantially to the total order parameter. Their magnitudes, as quantified in Table~\ref{tab:order}, are by no means negligible and must be taken into account for an accurate description of the superconducting state, as ARPES experiments on Bi2212 \cite{PhysRevLett.83.840} demonstrate that the observed gap anisotropy in the underdoped regime cannot be fitted without introducing these higher-order harmonic components.

\section{Appendix C: Determination of the Spin Correlation Length}

The real-space spin correlation function \(\chi(\mathbf{r})\) is obtained via Fourier transform of the static spin structure factor \(\chi(\mathbf{q})\) computed from the covariance+spin-GW solutions. For a square lattice of size \(N\times N\), we extract the correlation along the \(y\)-axis, \(\chi(0,y)\), to determine the spin correlation length \(\xi\). In the paramagnetic regime, the spatial decay at large distances follows an exponential form

\begin{equation}
|\chi(0,y)| \propto e^{-y/\xi}, \qquad y \gg 0,
\end{equation}

so that \(\xi\) can be extracted from a linear fit of \(\log|\chi(0,y)|\) versus \(y\):

\begin{equation}
\log|\chi(0,y)| = -\frac{1}{\xi}\,y + \text{const}.
\end{equation}

To avoid boundary effects and minimize finite-size artifacts, we restrict the fitting window to the intermediate range \(y\in[L/8,L/4]\), where the exponential decay is expected to be well behaved. This window yields stable and robust estimates of \(\xi\) across different temperatures.

\begin{figure}[htb]
  \centering
  \includegraphics[width=\textwidth]{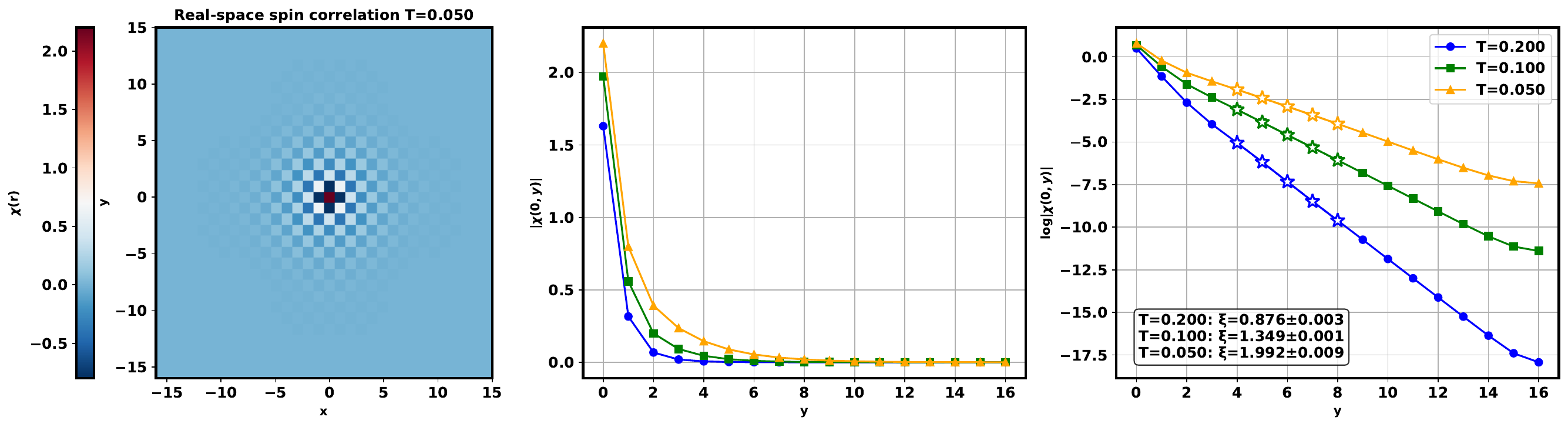}
  \caption{ Left panel: Real-space spin correlation map $\chi(\mathbf{r})$ on the $32\times32$ lattice.
  Middle panel: Absolute value $|\chi(0,y)|$ along the $y$-axis as a function of distance $y$.
  Right panel: Logarithmic plot $\log|\chi(0,y)|$ versus $y$; dashed lines denote linear fits performed in the fitting window $y\in[L/8,L/4]$ (with $L=32$), and the star symbols mark the data points used in the fits.
  All panels correspond to $U=8$, $t'=-0.25$, electron filling $n=0.93$, and lattice size $N=32\times32$.}
  \label{fig:spin_corr}
\end{figure}

Figure~\ref{fig:spin_corr} presents a representative case with \(U=8\), \(t'=-0.25\), \(N=32\times32\), and filling \(\langle n\rangle=0.93\). The extracted correlation lengths show a systematic increase upon cooling, consistent with the growth of magnetic fluctuations towards the antiferromagnetic instability. The linearity of the logarithmic decay confirms the exponential nature of the spin correlations in the studied temperature range.

\section{Appendix D: 2-Nambu spin-GW equation in the paramagnetic state}

We employ the four-Nambu representation to reformulate the action. 

\begin{equation}
\begin{aligned}
S = &-\sum_{AB}\int d(1,2)\Psi_A(1)\underline T^{AB}(1,2)\Psi_B(2) \\
&+ \frac{1}{2}\sum_{ab}\int d(1,2)\underline S_a(1)V^{ab}(1,2)\underline S_b(2).
\end{aligned}
\label{eq:action}
\end{equation}

Here $\Psi$ is the four-Nambu Grassmannian field. 
The four-component Nambu spinor $\Psi_{A}(1)$ ($A=1,2,3,4$) is defined as:
\begin{equation}
\Psi(1) \equiv 
\begin{pmatrix}
\psi_{1}(1) \\
\psi_{2}(1) \\
\psi_{3}(1) \\
\psi_{4}(1)
\end{pmatrix}
=
\begin{pmatrix}
\psi^*_{\uparrow}(1) \\
\psi_{\downarrow}(1) \\
\psi_{\uparrow}(1)\\
\psi^*_{\downarrow}(1)
\end{pmatrix}.
\label{eq:nambu_spinor}
\end{equation}

\(\underline{T}(1,2)\) represents the free part of the action:
\begin{equation}
\underline T(1,2)=\frac{1}{2}
\begin{pmatrix}
0 & T(1,2) \\
-T(2,1) & 0
\end{pmatrix},
\label{eq:T_matrix}
\end{equation}

And
\begin{equation}
T(1,2)=
\begin{pmatrix}
T_{\uparrow\uparrow}(1,2) & 0 \\
0 & -T_{\downarrow\downarrow}(2,1)
\end{pmatrix}.
\label{eq:T_components}
\end{equation}

with $a=(x,y,z)$, and 
$\underline S_a(1)=\sum_{AB}\Psi_A(1)\underline\sigma^a_{AB}\Psi_B(1)$, 
where $\underline\sigma^a$ are $4\times4$ antisymmetric matrices:
\begin{equation}
\begin{aligned}
\underline\sigma^x &= \frac{1}{2}
\begin{pmatrix}
0 & -1 & 0 & 0 \\
1 & 0 & 0 & 0 \\
0 & 0 & 0 & 1 \\
0 & 0 & -1 & 0
\end{pmatrix}
=\frac{i}{2}
\begin{pmatrix}
-\sigma^y & 0 \\
0 & \sigma^y
\end{pmatrix},\\
\underline\sigma^y &= \frac{1}{2}
\begin{pmatrix}
0 & -i & 0 & 0 \\
i & 0 & 0 & 0 \\
0 & 0 & 0 & -i \\
0 & 0 & i & 0 
\end{pmatrix}
=\frac{1}{2}
\begin{pmatrix}
\sigma^y & 0 \\
0 & \sigma^y
\end{pmatrix},\\
\underline\sigma^z &= \frac{1}{2}
\begin{pmatrix}
0 & 0 & -1 & 0 \\
0 & 0 & 0 & -1 \\
1 & 0 & 0 & 0 \\
0 & 1 & 0 & 0
\end{pmatrix}
=\frac{1}{2}
\begin{pmatrix}
0 & -\sigma^0 \\
\sigma^0 & 0
\end{pmatrix},
\end{aligned}
\label{eq:sigma_matrices}
\end{equation}
where $\sigma^0$ denotes the $2\times2$ identity matrix.

To derive the spin-GW equation, we need to introduce an external source \(J^a\) coupled to \(S_a\). At this point, the full action includes:
\begin{equation}
\begin{aligned}
S[J] = &-\sum_{AB}\int d(1,2)\Psi_A(1)\underline T^{AB}(1,2)\Psi_B(2) \\
&+ \frac{1}{2}\sum_{ab}\int d(1,2)\underline S_a(1)V^{ab}(1,2)\underline S_b(2)-\sum_{a}J^a(1) S_a(1).
\end{aligned}
\label{eq:actionJ}
\end{equation}

We follow the lowest-order Dyson-Schwinger equation~\cite{ROBERTS1994477}:
\begin{equation}
\int D[\Psi]\frac{\delta[\Psi_B(2)e^{-S}]}{\delta \Psi_A(1)}=0.
\label{eq:ds_eq}
\end{equation}

We obtain:
\begin{equation}
\begin{aligned}
\delta(1,2)\delta^{AB}
=&2\sum_C\int d(3)\,T^{AC}(1,3;J)\langle\Psi_B(2)\Psi_C(3)\rangle \\
&+2\sum_{ac}\sum_{CDE}\int d(3)\,\underline\sigma^{a}_{AC}
\langle\Psi_C(1)\Psi_D(3)\Psi_E(3)\Psi_B(2)\rangle \\
&\qquad\times V^{ac}(1,3)\underline\sigma^{c}_{DE}.
\end{aligned}
\label{eq:ds_result}
\end{equation}

where 
\begin{equation}
T(1,2;J)=T(1,2)+\sum_a J^a(1)\delta(1,2)\underline\sigma^a
\end{equation}

We define the off-shell four-Nambu Green's function:
\begin{equation}
\underline G_{AB}(1,2;J)=\langle\Psi_B(2)\Psi_A(1)\rangle.
\label{eq:green_function}
\end{equation}

In order to rewrite the two‑body correlation function, we define the 4-Nambu vertex function:
\begin{equation}
\Lambda^{a}_{AB}(1,2,3;J)=\frac{\underline G_{AB}(1,2;J)}{\delta J^a(3)}=\langle\Psi_{B}(2)\Psi_{A}(1)S_a(3)\rangle-\langle\Psi_{B}(2)\Psi_{A}(1)\rangle\langle S_a(3)\rangle.
\end{equation}

We can define the off‑shell Hartree self‑energy and the effective potential:
\begin{equation}
\begin{aligned}
    H^{-1}_{AB}(1,2;J)=&2T^{AB}(1,2;J)+\sum_a v^a(1;J)\underline\sigma^a_{AB}\delta(1,2),\\
    v^a(1;J) =&\, -2\sum_c\int d(3)\underline\sigma^c\underline G(3,3;J)V^{ac}(1,3).
\end{aligned}
\end{equation}

And We have the following identity:
\begin{equation}
\begin{aligned}
\Lambda^a(1,2,3;J)=&-\int d(4,5)\underline G(1,4;J)\frac{\delta G^{-1}(4,5;J)}{\delta J^a(3)}\underline G(5,2;J)\\
=&\,-\sum_{b}\int d(4,5,6)\underline G(1,4;J)\frac{\delta v^b(6;J)}{\delta J^a(3)}\frac{\delta G^{-1}(4,5;J)}{\delta v^b(6;J)}\underline G(5,2;J).
\end{aligned}
\end{equation}

At the lowest order of the vertex approximation:
\begin{equation}
\frac{\delta \underline G(1,2;J)}{\delta v^a(3;J)}\approx\frac{\delta H^{-1}(1,2;J)}{\delta v^a(3;J)}=\underline \sigma^a\delta(1,2)\delta(1,3).
\end{equation}

By adopting the on‑shell condition, we derive the set of self‑consistent equations for the 4‑Nambu spin‑GW:
\begin{equation}
\begin{aligned}
\underline G^{-1}_{AB}(1,2)
=& H^{-1}_{AB}(1,2) + 4\sum_{ac}\sum_{CD}
\underline\sigma^a_{AC}\,\underline G_{CD}(1,2)\,W^{ca}(2,1)\,\underline\sigma^c_{DB},\\
(V^{-1})^{ab}(1,2) =&\, (W^{-1})^{ab}(1,2) + 2P^{ab}(1,2),\\
P^{ba}(1,2) =&\, \operatorname{Tr}\big(\underline G(1,2)\,\underline\sigma^a
\,\underline G(2,1)\,\underline\sigma^b\big),\\
H^{-1}_{AB}(1,2) =&\, 2T^{AB}(1,2) + \sum_a v^a(1)
\,\underline\sigma^a_{AB}\,\delta(1,2),\\
v^a(1) =&\, -2\sum_{cDE}\int d(3)\,\underline\sigma_{DE}^{c}
\,\underline G^{ED}(3,3)\,V^{ac}(1,3).
\end{aligned}
\label{eq:gw_equations}
\end{equation}

For the Hubbard interaction,$V^{ab}(1,2)=-\frac{U}{3}\delta(a,b)\delta(1,2)$.
In the paramagnetic phase, the spin SU(2) symmetry remains unbroken. Therefore,
\begin{equation}
\underline G(1,2)=
\begin{pmatrix}
0 & -\tilde G^T(1,2) \\
\tilde G(1,2) & 0
\end{pmatrix}
\label{eq:G_structure}
\end{equation}

When superconducting symmetry breaking is permitted, $\tilde G$ reads
\begin{equation}
\tilde G(1,2)=
\begin{pmatrix}
G_{\uparrow\uparrow}(1,2) & F(1,2) \\
F^*(1,2) & -G_{\downarrow\downarrow}(2,1)
\end{pmatrix},
\label{eq:tilde_G}
\end{equation}
and the transpose of $\tilde G$ is
\begin{equation}
\tilde G^T(1,2)=
\begin{pmatrix}
G_{\uparrow\uparrow}(2,1) & F^*(2,1) \\
F(2,1) & -G_{\downarrow\downarrow}(1,2)
\end{pmatrix},
\label{eq:tilde_G_transpose}
\end{equation}

where $G_{\alpha\alpha}(1,2)=\langle\psi_{\alpha}(2)\psi^*_{\alpha}(1)\rangle$ is the normal Green's function, 
and $F(1,2)=\langle \psi_{\downarrow}(2)\psi_{\uparrow}(1)\rangle$, 
$F^*(1,2)=\langle \psi^*_{\uparrow}(2)\psi^*_{\downarrow}(1)\rangle$ are the anomalous components.

The above relations can decouple the 4‑Nambu equations into two equivalent 2‑Nambu equations:
\begin{equation}
\begin{aligned}
\tilde G^{-1}(1,2) 
=& T(1,2) + \frac{U}{3}\operatorname{Tr}\big(\tilde G(1,1)\big)\delta(1,2)
- 2\sigma_y\tilde G^T(1,2)\sigma_y W^{xx}(2,1)\\
+& \tilde G(1,2) W^{zz}(2,1) + 2i\sigma_y\tilde G^T(1,2)\sigma_y W^{yx}(2,1),\\
(V^{-1})^{ab}(1,2) =&\, (W^{-1})^{ab}(1,2) + 2P^{ab}(1,2).
\end{aligned}
\label{eq:gw_2x2}
\end{equation}

For the Lindhard component, because the SU(2) symmetry is still preserved under the vertex truncation,
we adopt the definitions:
\begin{equation}
\begin{aligned}
P^{zz}(1,2) 
=& \frac{1}{4}\operatorname{Tr}\Big[\tilde G(1,2)\tilde G(2,1) 
+ \tilde G^{T}(1,2)\tilde G^{T}(2,1)\Big] \\
=& \frac{1}{2}\operatorname{Tr}\Big[\tilde G(1,2)\tilde G(2,1)\Big],\\
P^{xx}(1,2) =&\, P^{yy}(1,2) = P^{zz}(1,2),\\
P^{ab}(1,2) =&\, 0 \quad (a \ne b).
\end{aligned}
\label{eq:lindhard}
\end{equation}

We define the Fourier transform:
\begin{equation}
\begin{aligned}
\tilde G(1,2) &= \tilde G(1-2) 
= \frac{1}{\mathcal{N}}\int dk\,\tilde G(k)e^{ik(1-2)},\\
\tilde G(k) &= \int d(1-2)\,\tilde G(1)e^{-ik(1-2)}.
\end{aligned}
\label{eq:fourier}
\end{equation}

Finally, the set of equations for the 2‑Nambu spin‑GW in momentum‑frequency space is derived.
\begin{equation}
\begin{aligned}
\tilde G^{-1}(k) &= T(k) + \frac{3}{\mathcal{N}}\sum_{q}\tilde G(k+q)W(q),\\
W^{-1}(q) &= -\frac{3}{U}- 2P(q),\\
P(q) &= \frac{1}{2\mathcal{N}}\sum_{k}\operatorname{Tr}\big[\tilde G(k+q)\tilde G(k)\big].
\end{aligned}
\label{eq:gw_momentum}
\end{equation}

% ----- 附录 C -----
\section{Appendix E: Covariance calculation in the normal phase and superconducting phase}
%\input{appendix/appendixC.tex}
% \section{ Covariance Calculation in the normal phase and superconducting phase}
% \label{sec:dwave_correlation}

In the two spinor representation, we introduce an external source $\phi$, which couples to the order parameter $X$ and thus contributes an action:
\begin{equation}
    S^{\text{ext}}=-\int d(1)\phi(1)X(1).
\end{equation}

In general, the order parameter $X(1)$ has the following form:
\begin{equation}
    X(1)=\sum_{ab}\int d(2,3)\Psi^a(2) K_X^{ab}(1,2,3)\Psi^b(3),
\end{equation}
where $K_X(1,2,3)$ is the kernel associated with the order parameter. For the d-wave order parameter,
\begin{equation}
    K_X(1,2,3)=
    \begin{pmatrix}
        0 & \delta(1,2)\kappa(2,3) \\
        0 & 0
    \end{pmatrix},
\end{equation}
with
\begin{equation}
\begin{aligned}
    \kappa(1,2)=\delta(\tau_1,\tau_2)\bigl[
    &\delta_{\mathbf{x}_1,\mathbf{x}_1+\mathbf{e}_x}
    +\delta_{\mathbf{x}_1,\mathbf{x}_1-\mathbf{e}_x} \\
    &-\delta_{\mathbf{x}_1,\mathbf{x}_1+\mathbf{e}_y}
    -\delta_{\mathbf{x}_1,\mathbf{x}_1-\mathbf{e}_y}\bigr].
\end{aligned}
\end{equation}

We can absorb the external source term into the free term, obtaining a free term that depends on the external source:
\begin{equation}
    T_{\phi}(1,2)=
    \begin{pmatrix}
        T_{\uparrow}(1,2) & \phi(1)\kappa(1,2) \\
        0 & -T_{\downarrow}(2,1)
    \end{pmatrix}.
\end{equation}

Next, we derive the covariant equation based on the spin-GW approach. Without loss of generality, we first consider an external source of arbitrary form that couples to a two-fermion term, forming a new effective free term:
\begin{equation}
    \begin{aligned}
    \tilde \Gamma_{\phi}(1,2,3)
    =& \tilde \gamma_{\phi}(1,2,3)
    +\frac{U}{3}\operatorname{Tr}\big(\tilde \Lambda_{\phi}(1,1,3)\big)\delta(1,2) \\
    &\quad + \tilde \Lambda_{\phi}(1,2,3)W(2,1)
    +\tilde G(1,2)\Lambda_{w}(2,1,3),\\
    \Lambda_{w}(1,2,3) =& -\int d(4,5)W(1,4)\Gamma_{w}(4,5,3)W(5,2),\\
    \Gamma_{w}(1,2,3)=&
    -\operatorname{Tr}\bigl[
    \tilde \Lambda_{\phi}(1,2,3)\tilde G(2,1)
    +\tilde G(1,2)\tilde \Lambda_{\phi}(2,1,3)\bigr],\\
     \tilde \Lambda_{\phi}(1,2,3)=&
    -\int d(4,5)\tilde G(1,4)\tilde \Gamma_{\phi}(4,5,3)\tilde G(5,2).
    \end{aligned}
\end{equation}

The vertex functions are defined as
\begin{align}
    \tilde \Gamma_{\phi}(1,2,3) &= \frac{\delta \tilde G^{-1}_{\phi}(1,2)}{\delta \phi(3)}, \\
    \tilde \Lambda_{\phi}(1,2,3) &= \frac{\delta \tilde G_{\phi}(1,2)}{\delta \phi(3)}, \\
    \tilde \Gamma_{w}(1,2,3) &= \frac{\delta W^{-1}(1,2)}{\delta \phi(3)}, \\
    \Lambda_{w}(1,2,3) &= \frac{\delta W(1,2)}{\delta \phi(3)}.
\end{align}

Assuming the external source couples to the d-wave order parameter and the system is in the paramagnetic state, it is easy to verify that:
\begin{equation}
    \Lambda_{w}(1,2,3)=0, \quad
    \operatorname{Tr}\big(\tilde \Lambda_{\phi}(1,1,3)\big)=0.
\end{equation}

Moreover, we can reduce the two-component covariant equation to a single-component equation:
\begin{equation}
    \begin{aligned}
    \Gamma_{\phi}(1,2,3)
    &= \kappa(1,2)\delta(1,3)
    +3\Lambda_{\phi}(1,2,3)W(2,1), \\
    \Lambda_{\phi}(1,2,3)
    &= -\int d(4,5) G(4,1) \Gamma_{\phi}(4,5,3) G(5,2).
    \end{aligned}
\end{equation}

Since the vertex function has three space-time indices, we assume that the system possesses translational invariance to reduce the number of indices. In this case,
\begin{equation}
    f(1,2,3)=f(1-2,1-3).
\end{equation}

We define the Fourier transform and its inverse for three-index space-time quantities:
\begin{equation}
    \begin{aligned}
    f(1,2,3)
    &= \frac{1}{\mathcal{N}^2}\sum_{k,q}f(k,q)e^{ik(1-2)+iq(1-3)}, \\
    f(k,q)
    &= \int d(1-2,1-3)f(1,2,3)e^{-ik(1-2)-iq(1-3)}.
    \end{aligned}
\end{equation}

For the covariant equation that couples to the d-wave order parameter, we express the equations in momentum-frequency space:
\begin{equation}
    \begin{aligned}
    \Gamma_{\phi}^d(k;q) &= \kappa(k)+\Gamma_{\text{MT}}^d(k;q), \\
    \Gamma_{\text{MT}}^d(k;q) &= \frac{3}{\mathcal{N}}\sum_{p}\Lambda_{\phi}^d(k+p;q)W(p), \\
    \Lambda_{\phi}^d(k;q) &= G(-k-q)\Gamma_{\phi}^d(k;q)G(k).
    \end{aligned}
\end{equation}

If we add an action that couples to the density-type operator
\( n(1) = \psi^*_{\uparrow}(1)\psi_{\uparrow}(1) + \psi^*_{\downarrow}(1)\psi_{\downarrow}(1) \)
and the \(z\)-direction spin operator
\( S^z(1) = \psi^*_{\uparrow}(1)\psi_{\uparrow}(1) - \psi^*_{\downarrow}(1)\psi_{\downarrow}(1) \),
the spin-singlet structure of the system remains unchanged. That is, the 4-Nambu Green's function only has four components:
\( \langle \psi_{\uparrow}^*(2)\psi_{\uparrow}(1) \rangle \),
\( \langle \psi_{\downarrow}^*(2)\psi_{\downarrow}(1) \rangle \),
\( \langle \psi_{\uparrow}(2)\psi_{\downarrow}(1) \rangle \),
and \( \langle \psi_{\downarrow}(2)\psi_{\uparrow}(1) \rangle \).
Consequently, we can still retain the spin-GW form of Eq.~(\ref{eq:gw_equations}).

Performing the functional derivative of Eq.~(\ref{eq:gw_equations}), we obtain:
\begin{equation}
\begin{aligned}
\Gamma_{\phi}(1,2,3) &=\gamma_{\phi}(1,2,3)+\frac{U}{3}\operatorname{Tr}\bigl(\dot{\tilde G}_{\phi}(1,1,3)\bigr)\delta(1,2) \\
&\quad -2\sigma^y \dot{\tilde G}_{\phi}^T(1,2,3)\sigma^y W_{\phi}^{xx}(2,1) \\
&\quad -2\sigma^{y}\tilde G^T(1,2)\sigma^y \Lambda_{w,\phi}^{xx}(2,1,3) \\
&\quad +\dot{\tilde G}_{\phi}(1,2,3) W_{\phi}^{zz}(2,1) \\
&\quad +\tilde{G}(1,2) \Lambda_{w,\phi}^{zz}(2,1,3) \\
&\quad +2i\sigma^y\dot{\tilde G}_{\phi}^T(1,2,3)\sigma^y W_{\phi}^{yx}(2,1) \\
&\quad +2i\sigma^y\tilde G^T(1,2)\sigma^y \Lambda_{w,\phi}^{yx}(2,1,3),
\end{aligned}
\end{equation}
which defines the derivative of the vertex function.

The inverse derivative of the screened interaction and the three components of $\dot{P}_{\phi}^{ab}$ are given by:
\begin{equation}
\begin{aligned}
(\Lambda_{w,\phi}^{-1})^{ab}(1,2,3) &= -2 \dot{P}_{\phi}^{ab}(1,2,3), \\[4pt]
\dot{P}_{\phi}^{xx}(1,2,3) &= -\frac{1}{2}\operatorname{Tr}\Bigl(\sigma^y\tilde G(1,2)\sigma^y\dot{\tilde G}_{\phi}(1,2,3) \\
&\qquad +\sigma^y\tilde G(2,1)\sigma^y\dot{\tilde G}_{\phi}(2,1,3)\Bigr), \\[4pt]
\dot{P}_{\phi}^{zz}(1,2,3) &= \frac{1}{2}\operatorname{Tr}\Bigl(\tilde G(1,2)\dot{\tilde G}_{\phi}(2,1,3) \\
&\qquad +\tilde G(2,1)\dot{\tilde G}_{\phi}(1,2,3)\Bigr), \\[4pt]
\dot{P}_{\phi}^{xy}(1,2,3) &= \frac{i}{2}\operatorname{Tr}\Bigl(\sigma^y\tilde G(2,1)\sigma^y\dot{\tilde G}_{\phi}(2,1,3) \\
&\qquad -\sigma^y\tilde G(1,2)\sigma^y\dot{\tilde G}_{\phi}(1,2,3)\Bigr),
\end{aligned}
\end{equation}
where the superscripts label the spin components.

Finally, the functional derivatives of the screened interaction and the Green's function read:
\begin{equation}
\begin{aligned}
\Lambda_{w,\phi}^{ab}(1,2,3) &= -\sum_{cd}\int d(4,5)\, W_{\phi}^{ac}(1,4) \\
&\qquad \times (\Lambda_{w,\phi}^{-1})^{cd}(4,5,3) W_{\phi}^{db}(5,2), \\[4pt]
\dot{\tilde G}_{\phi}(1,2,3) &= -\int d(4,5)\,\tilde G(1,4)\,\dot{\tilde G}_{\phi}^{-1}(4,5,3)\,\tilde G(5,2).
\end{aligned}
\label{eq:gw_covariance}
\end{equation}

In the superconducting solution, since
\( W^{ab}(1,2) = W(1,2) \delta^{ab} \),
\( \tilde G^T(1,2) = \tilde G(2,1) \),
and \( \dot{\tilde G}_{\phi}^T(1,2,3) = \dot{\tilde G}_{\phi}(2,1,3) \),
we have the self-consistent equations in momentum-frequency space:

\begin{equation}
\begin{aligned}
\Gamma_{\phi}(k,q) &=\gamma_{\phi}(k,q)+\frac{U}{3\mathcal{N}}\sum_{k'} \operatorname{Tr}\bigl(\dot{\tilde G}_{\phi}(k',q)\bigr) \\
&\quad -\frac{2}{\mathcal{N}}\sum_{k'}\sigma^{y}\dot{\tilde G}_{\phi}(-k-k'-q,q)\sigma^y W_{\phi}^{xx}(k') \\
&\quad -\frac{2}{\mathcal{N}}\sum_{k'}\sigma^{y}\Lambda_{w,\phi}^{xx}(-k-k'-q,q)\sigma^y \tilde G(k') \\
&\quad +\frac{1}{\mathcal{N}}\sum_{k'}\dot{\tilde G}_{\phi}(k+k',q) W_{\phi}^{zz}(k') \\
&\quad +\frac{1}{\mathcal{N}}\sum_{k'}\Lambda_{w,\phi}^{zz}(k'-q-k,q)\tilde G(k') \\
&\quad +\frac{2i}{\mathcal{N}}\sum_{k'}\sigma^y\Lambda_{w,\phi}^{yx}(-k-k'-q,q)\sigma^y\tilde G(k'),
\end{aligned}
\end{equation}
which defines the vertex function in momentum-frequency space.

In the same representation, the inverse derivative of the screened interaction and the three components of $\dot{P}_{\phi}^{ab}$ are:
\begin{equation}
\begin{aligned}
(\Lambda_{w,\phi}^{-1})^{ab}(k,q) &= -2 \dot{P}_{\phi}^{ab}(k,q), \\[4pt]
\dot{P}_{\phi}^{xx}(k,q) &= -\frac{1}{\mathcal{N}}\sum_{k'}\operatorname{Tr}\Bigl(\sigma^y \tilde G(k')\sigma^y\dot{\tilde G}_{\phi}(k-k',q) \\
&\qquad +\sigma^y\tilde G(k')\sigma^y\dot{\tilde G}_{\phi}(-k-q-k',q)\Bigr), \\[4pt]
\dot{P}_{\phi}^{zz}(k,q) &= \frac{1}{\mathcal{N}}\sum_{k'}\operatorname{Tr}\Bigl(\dot{\tilde G}_{\phi}(k+k',q)\tilde G(k') \\
&\qquad +\dot{\tilde G}_{\phi}(k'-q-k,q)\tilde G(k')\Bigr), \\[4pt]
\dot{P}_{\phi}^{xy}(k,q) &= \frac{i}{\mathcal{N}}\sum_{k'}\operatorname{Tr}\Bigl(-\sigma^y \tilde G(k')\sigma^y\dot{\tilde G}_{\phi}(k-k',q) \\
&\qquad +\sigma^y\tilde G(k')\sigma^y\dot{\tilde G}_{\phi}(-k-q-k',q)\Bigr),
\end{aligned}
\end{equation}
where the superscripts denote spin components.

Finally, the momentum-frequency expressions for the derivatives of the screened interaction and the Green's function are:
\begin{equation}
\begin{aligned}
\Lambda_{w,\phi}^{xx}(k,q) &= -W(k+q)\,(\Lambda_{w,\phi}^{-1})^{xx}(k,q)\,W(k), \\[4pt]
\Lambda_{w,\phi}^{zz}(k,q) &= -W(k+q)\,(\Lambda_{w,\phi}^{-1})^{zz}(k,q)\,W(k), \\[4pt]
\Lambda_{w,\phi}^{xy}(k,q) &= -W(k+q)\,(\Lambda_{w,\phi}^{-1})^{xy}(k,q)\,W(k), \\[4pt]
\dot{\tilde G}_{\phi}(k,q) &= -G(k+q)\,\Gamma_{\phi}(k,q)\,G(k).
\end{aligned}
\label{eq:gw_covariance_mom}
\end{equation}

To compute $\chi^{zz}(q)$, we have:
\begin{equation}
\gamma_{\phi}(k,q) = \sigma^0, \qquad 
\chi^{zz}(q) = \sum_k \operatorname{Tr}\bigl(\dot{\tilde G}_{\phi}(k,q)\bigr).
\end{equation}

Similarly, to compute $\chi^{00}(q)$, we have:
\begin{equation}
\gamma_{\phi}(k,q) = \sigma^z, \qquad 
\chi^{00}(q) = \sum_k \operatorname{Tr}\bigl(\sigma^z\dot{\tilde G}_{\phi}(k,q)\bigr).
\end{equation}

\section{Appendix F: Details of the finite-size scaling analysis}
%\input{appendix/appendixD.tex}
% \section{Details of the finite-size scaling analysis}
The momentum-space correlation function near a continuous phase transition obeys the universal scaling form~\cite{Georgii}
\begin{equation}
\chi(\boldsymbol{q}, T) = \frac{\chi_0(T)}{1 + (\xi(T) q)^{2}},
\end{equation}
where the correlation length diverges as~\cite{Fisher_1967}
\begin{equation}
\xi(T) = \xi_0 |T - T_c(\infty)|^{-\nu},
\end{equation}
with \(T_c(\infty)\) the critical temperature in the thermodynamic limit \(L\to\infty\), and \(\nu\) the correlation length exponent. The zero-momentum correlation function satisfies
\begin{equation}
\chi_0(T) = \chi(\boldsymbol{q}=\mathbf{0}, T) \sim |T - T_c(\infty)|^{-\gamma},
\end{equation}
with \(\gamma\) the susceptibility exponent.

The real-space correlation function \(\chi(\boldsymbol{r}, T)\), obtained via inverse Fourier transform of \(\chi(\boldsymbol{q}, T)\), has the closed-form solution
\begin{equation}
\chi(r, T) \sim K_0\left(\frac{r}{\xi(T)}\right),
\label{eq:k0}
\end{equation}
\(K_0(x)\) is the zeroth-order modified Bessel function of the second kind

For finite systems, we often define the transition temperature as the temperature corresponding to the correlation length around one system size \( L \).
\begin{equation}
    \xi(T_c)\sim L
\end{equation}
According to Eq.~(\ref{eq:k0}), if the correlation length is fixed, the ratio of the correlation functions at any two points in space is constant. 
For convenience, we take the ratio of the correlation functions corresponding to \(L/2\) and \(L/4\) here.
This criterion is mathematically expressed as
\begin{equation}
\frac{\chi( L/2,T_c)}{\chi( L/4,T_c)} = C,
\end{equation}
where \(C\) is a universal constant independent of \(L\). The temperature \(T\) that satisfies this equation is defined as the finite-size critical temperature \(T_c(L)\). 
This method effectively eliminates the influence of finite-size effects on the critical temperature determination, 
providing a reliable approach to extract \(T_c(L)\) for different system sizes, 
which is further used to extrapolate \(T_c(\infty)\) via the scaling relation
\begin{equation}
 T_c(L) - T_c(\infty) \sim \beta_c(\infty)-\beta_c(L)\sim L^{-1/\nu}.
\end{equation}

\begin{figure}[htb]
  \centering
  \includegraphics[width=0.6\textwidth]{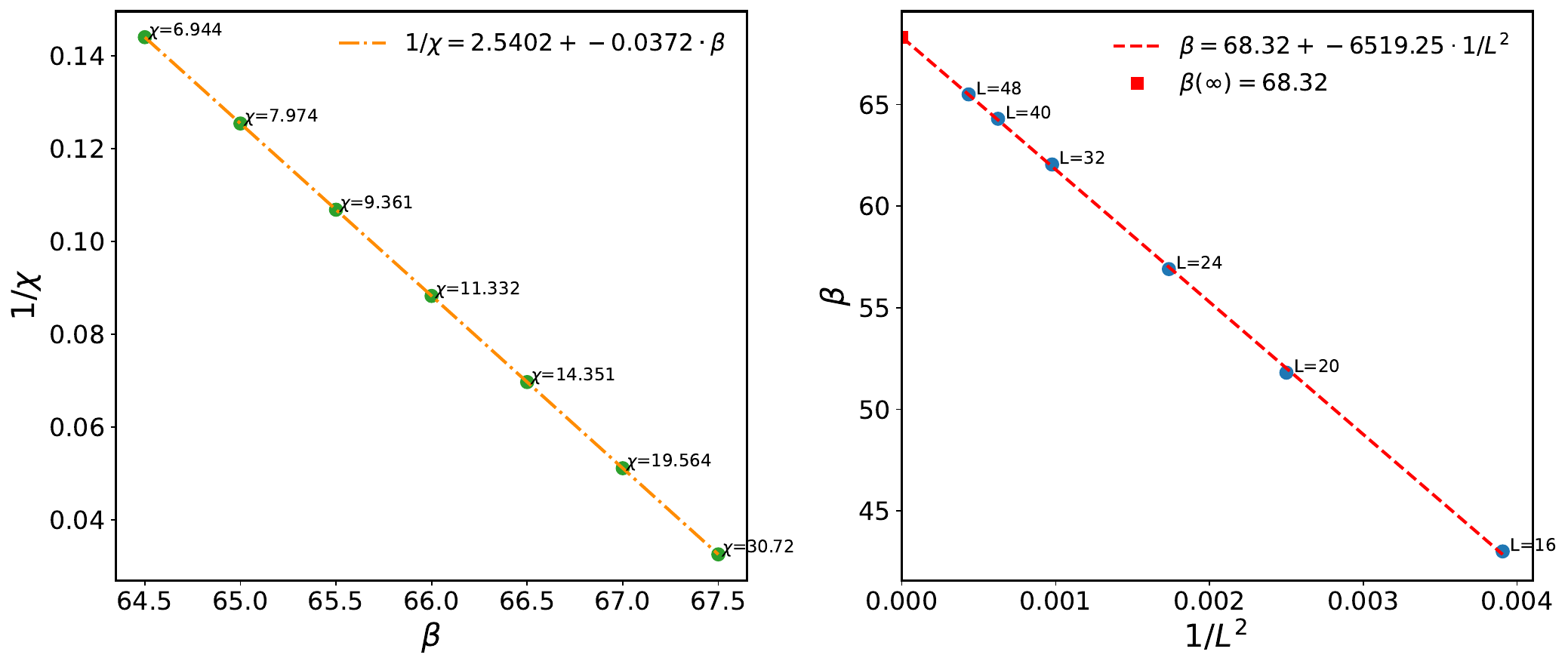} 
  \caption{Left: $U=8$, $t'=-0.25$, $N=32\times32$, $\langle n\rangle=0.822$. Temperature dependence of the $d$-wave superconducting correlation function at zero Matsubara frequency and zero momentum. Right: Same parameters as left panel. $\beta_c(L)$ is determined by the ratio $\chi_{L/2,T}/\chi_{L/4,T}$, and $\beta_c(\infty)$ denotes the $d$-wave superconducting transition temperature in the thermodynamic limit.}
  \label{fig:Figure6}
\end{figure}

It can be seen from Fig.~\ref{fig:Figure6} that the critical exponents for this phase transition are $\gamma=1$ and $\nu=1/2$, which are consistent with the characteristics of mean-field phase transitions. This confirms that the transition behavior of the $d$-wave superconductivity obtained by our method is BCS-like.

% ---------- 补充材料可选的参考文献（如有需要） ----------
% \bibliography{apssamp}

% \end{document}
% \bibliographystylesupp{apsrev4-2}
% ---- 附录参考文献列表（独立编号，带 S 前缀） ----
\end{document}